\documentclass[10pt,conference]{IEEEtran}
\usepackage{graphicx}

\begin{document}
\title{A low-resolution, GSa/s streaming digitizer for a correlation-based trigger system}
\author{\IEEEauthorblockN{Kurtis~Nishimura,
			Matthew~Andrew,
			Zhe~Cao,
			Michael~Cooney,\\
			Peter~Gorham,
			Luca~Macchiarulo,
			Lisa~Ritter,
			and~Gary~Varner}
	\IEEEauthorblockA{Department of Physics and Astronomy\\
			University of Hawai`i at M\={a}noa\\
			2505 Correa Road,\\
			Honolulu, HI 96822}
	\and
	\IEEEauthorblockN{Andr\'es~Romero-Wolf}
	\IEEEauthorblockA{Jet Propulsion Laboratory\\
			4800 Oak Grove Drive\\
			Pasadena, CA 91109}
}

\maketitle

\begin{abstract}
Searches for radio signatures of ultra-high energy neutrinos and cosmic rays 
could benefit from improved efficiency by using real-time beamforming or 
correlation triggering.  For missions with power limitations, such as the 
ANITA-3 Antarctic balloon experiment, full speed high resolution digitization 
of incoming signals is not practical.  To this end, the University of Hawaii 
has developed the  Realtime Independent Three-bit Converter (RITC), 
a 3-channel, 3-bit, streaming analog-to-digital converter implemented in the 
IBM-8RF 0.13~$\mu$m process.  
RITC is primarily designed to digitize broadband radio signals produced by the 
Askaryan effect, and thus targets an analog bandwidth of $>$1~GHz, with a 
sample-and-hold architecture capable of storing up to 
2.6~gigasamples-per-second.  An array of flash analog-to-digital converters 
perform 3-bit conversion of sets of stored samples while acquisition continues 
elsewhere in the sampling array. A serial interface is provided to access an 
array of on-chip digital-to-analog converters that control the digitization 
thresholds for each channel as well as the overall sampling rate. 
Demultiplexed conversion outputs are read out simultaneously for each channel 
via a set of 36 LVDS links, each running at 650~Mb/s. 
We briefly describe the design architecture of RITC.  Evaluation of the RITC
is currently under way, and we will report testing updates as they become 
available, including prospects for the use of this architecture as the analog 
half of a novel triggering system for the ANITA-3 ultra-high energy neutrino 
experiment.
\end{abstract}

\section{Introduction}
In two long-duration balloon flights over the Antarctic continent, the ANITA 
experiment has placed the most restictive limits to date on ultra-high energy 
neutrino flux \cite{ANITA1,ANITA2}, as well as observing a number of 
ultra-high energy cosmic ray events \cite{ANITA-UHECR}.
The trigger systems for these flights are based on dividing incoming antenna 
signals into frequency bands.  To search for 
broadband signals consistent with an impulsive event, power is measured in 
various bands, and a trigger is generated when power thresholds are exceeded 
in multiple bands for multiple adjacent sectors of antennas.  While this 
trigger has proved sufficient for the first two flights, it does not benefit 
from the full interferometric potential of the ANITA instrument.  

\begin{figure}
\centering
\includegraphics[width=0.9\columnwidth]{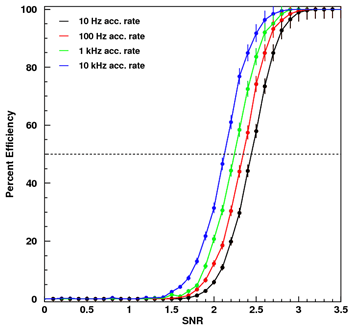}
\caption{RITC enabled simulated efficiency curves for various accidentals 
rates using a beamforming trigger scheme.  In this particular scheme, 
beamforming is performed on individual $\phi$-sectors (sets of three 
antennas arranged vertically on the payload), and total power of the 
beamforming output is combined between sets of three adjacent $\phi$-sectors.}
\label{fig:eff_curves}
\end{figure}

The proposed upgrade to the trigger system utilizes realtime interferometric 
techniques to search for correlated signals between antennas which can point 
back to a source region of interest.  A key component of this technique is 
the ability to digitize all incoming data so that such interferometry can be 
performed. In order to limit the power requirements of such a system, only 
low-resolution digitization is performed. This digitized data is then 
processed through digital logic to generate a system trigger, which results in
high resolution digitization of the event.  Simulated efficiency curves, 
shown in Figure \ref{fig:eff_curves}, have been generated for this triggering 
technique, and indicate that 50\% efficiency could be reached at an SNR of 
2.3 or below, compared to SNRs $>3.5$ obtained in the previous flights 
\cite{ANITA-INSTRUMENT,ANITA2}.

We have fabricated an ASIC, designated the Realtime Independent Three-bit 
Converter, in IBM 0.13~$\mu$m technology to evaluate the digitization aspect 
of this proposed trigger system.

\section{Architecture}

\begin{figure}
\centering
\includegraphics[width=1.0\columnwidth]{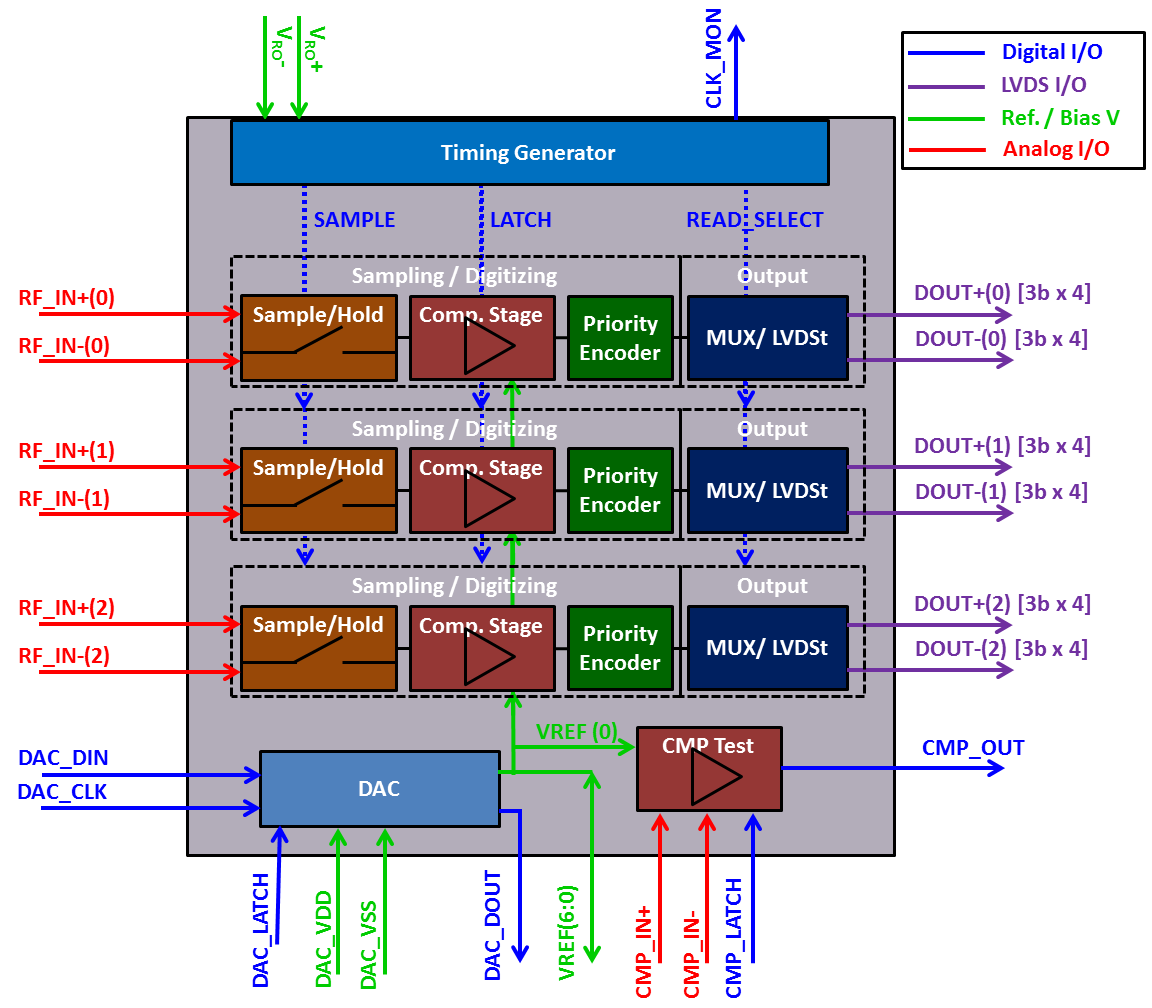}
\caption{Block diagram of RITC architecture.}
\label{fig:RITC_block}
\end{figure}

The basic RITC architecture is shown in block-diagram form in 
Figure \ref{fig:RITC_block}.  It consists of three differential analog input 
channels, each coupled into an array of 32~sample-and-hold cells.  
When an input clock pulse is sent from an external controlling device, 
such as an FPGA, this pulse traverses a delay line, and logic implemented at
variaous stages of the delay line dictates the track and hold periods for
each sampling cell.  After the hold period is entered, internal logic 
begins the digitization process, which is conducted in parallel for four 
sample cells on all three channels simultaneously.  The ADC is based on a 
flash architecture, with seven user-adjustable thresholds that define the 
digitization levels.  This allows flexibility to adjust for potential 
nonlinearities in the comparators, change the digitized levels in response 
to increased input noise levels, or utilize unequally spaced bit intervals, 
if desired.
The digitization thresholds, as well as biases that set the effective sampling
rate, are controlled by a set of internal DACs, which can be set by a serial 
programming interface.  Final digitized outputs are read out in parallel
sets of 36 bits (3 bits from 4 samples of 3 channels).  The outputs
are implemented as differential LVDS pairs, which must operate at 
650~Mb/s at the nominal sampling rate of 2.6~GSa/s.  

\begin{figure}
\centering
\includegraphics[width=0.9\columnwidth]{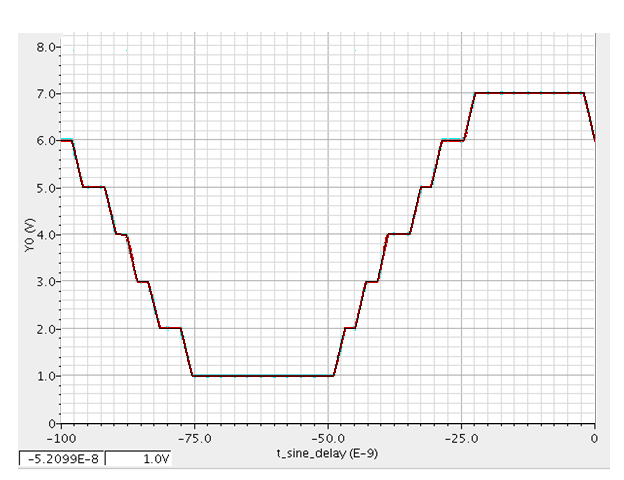}
\caption{A simulation of RITC three-bit digitization of a 20~MHz sine wave 
input.  The vertical axis is in ADC counts.}
\label{fig:RITC_sine_sim}
\end{figure}

Test and monitoring structures are also implemented.  A stand-alone comparator
is provided to allow characterization of linearity.  Analog outputs provide
access to the interal DAC settings for one of the three RITC channels.  
These allow evaluation of the DAC performance, as well as override capability
in the event of a DAC or shift register failure.  A reference clock output
is derived from the input clock, combined with the internal sampling logic,
to allow monitoring of the internal sampling rate.  This type of monitoring
also allows the user to implement feedback controls to compensate for 
temperature drifts.  A simulated output of the RITC response to a sine 
wave input is shown in Figure \ref{fig:RITC_sine_sim}.

\begin{figure}
\centering
\includegraphics[width=0.9\columnwidth]{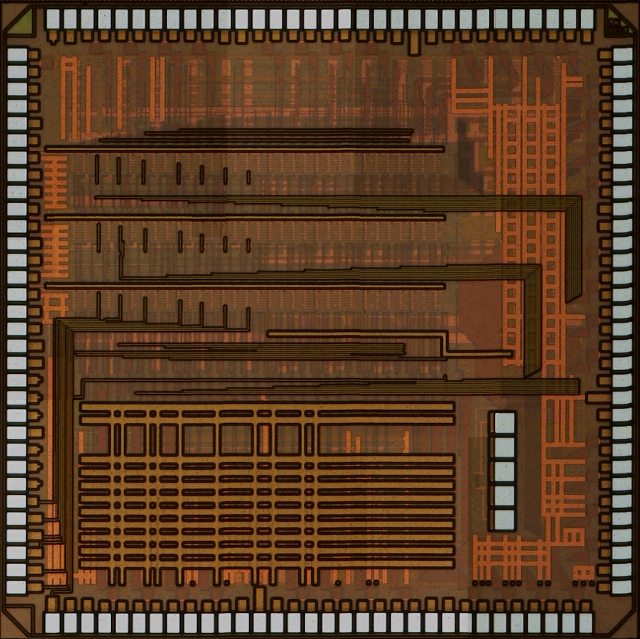}
\caption{Die photograph of RITC, as fabricated.  The die size is 3.13$\times$3.13~mm$^2$.}
\label{fig:RITC_die}
\end{figure}

RITC has been fabricated in through the MOSIS multi-project wafer service. 
A photograph of the bare die can be seen in Figure \ref{fig:RITC_die}.  Though
the design is not particularly space intensive, the large demultiplexing 
required to read out the data results in a pin-limited design of total 
dimension 9.8~mm$^2$.  

\section{Testing Status and Plans}
\begin{figure}
\centering
\includegraphics[width=0.9\columnwidth]{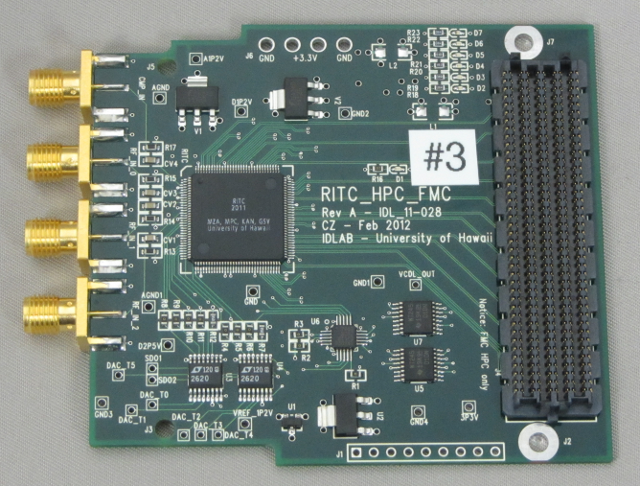}
\caption{A photograph of a fully populated FMC evaluation card for RITC.}
\label{fig:RITC_eval}
\end{figure}

We have designed a PCB incorporating the RITC to perform evaluation studies of
ASIC performance (see Figure \ref{fig:RITC_eval}). 
This board follows the FMC standard, and is 
intended for testing with the Xilinx ML605 evaluation platform.  This
interface will ultimately allow testing of RITC digitization with
beamforming implementations on either an FPGA or a recently submitted 
digital ASIC.
Testing is underway now, and we expect shortly to have evaluation results of
RITC performance, including analog bandwidth, comparator and DAC linearity, 
timing performance, power consumption, and overall suitability for use in 
the beamforming trigger system.

\section{Conclusion}
The RITC ASIC has been designed as a prototype of the analog functionality
required for a realtime interferometric trigger for the ANITA-3 experiment.  
A separate, fully digital ASIC consisting of the beamforming trigger logic 
has also been submitted for fabrication.
Once available, we will report test results of the RITC and the outlook for the
ANITA-3 trigger upgrade.  
Although targeted for ANITA-3, this type of trigger 
could enable significant improvements in noise-limited sensitivity for
next-generation radio searches for high energy neutrinos and cosmic rays.


\begin{thebibliography}{1}

\bibitem{ANITA1}
P.~Gorham {\it et al.}, Phys. Rev. Lett. {\bf 103}, 051103 (2009).

\bibitem{ANITA2}
P.~Gorham {\it et al.}, Phys. Rev. {\bf D82}, 022004 (2010).

\bibitem{ANITA-UHECR}
S.~Hoover {\it et al.}, Phys. Rev. Lett. {\bf 105}, 151101 (2010).

\bibitem{ANITA-INSTRUMENT}
P.~Gorham {\it et al.} Astropart. Phys. 32, 10 (2009).

\end{thebibliography}
\end{document}